\documentclass[mathleft]{an}
\usepackage{graphicx}
\usepackage{times}
\def\kms{km$\,$s$^{-\!1}$} 
\def\vsi{$v\: \sin i$}

\newcommand{\ltsima} {$\; \buildrel < \over \sim \;$} 
\newcommand{\simlt} {\lower.5ex\hbox{\ltsima}} % < over MMM 
\newcommand{\gtsima} {$\; \buildrel > \over \sim \;$} 
\newcommand{\simgt} {\lower.5ex\hbox{\gtsima}} % > over MMM 

\begin{document}

% The following seven commands are intended for editorial usage and should be ignored by
% the author(s).
\Pagespan{789}{}% Document's page range. 
% If second parameter is left empty, the last page is computed automatically.
\Yearpublication{2009}%
\Yearsubmission{2009}%
\Month{11}%   
\Volume{999}%  
\Issue{88}% 
% \DOI{This.is/not.aDOI}% 

\title{Orbital eccentricity of the symbiotic star MWC~560 
  \thanks{based on: observations obtained in ESO program 074.D-0114
  } }

\author{R. K. Zamanov\inst{1}\fnmsep\thanks{corresponding author:
  \email{rkz@astro.bas.bg}\newline}
%Example 
%for footnote, note the usage of the \texttt{fnmsep}
%command as separator between institute number and footnote mark} 
\and A. Gomboc\inst{2}
\and K. A. Stoyanov\inst{1}
\and I. K. Stateva\inst{1}
}

\titlerunning{MWC~560 : orbital eccentricity}
\authorrunning{Zamanov, Gomboc, Stoyanov, et al. }
\institute{Institute of Astronomy, Bulgarian Academy of Sciences, 
       72 Tsarigradsko Shousse Blvd., 1784 Sofia, Bulgaria
       \and 
     Faculty of Mathematics and Physics, University of Ljubljana, 
     Jadranska 19, 1000 Ljubljana, Slovenia  }

\received{30 June 2009}
\accepted{29 Oct  2009}
\publonline{later}


\abstract{%
 We present projected rotational velocity measurements of the red giant in the symbiotic 
 star MWC~560, using the high-resolution spectroscopic observations 
 with the FEROS spectrograph.
 We find that the projected rotational velocity of the red giant is \vsi $= 8.2 \pm 1.5$~\kms, 
 and estimate its rotational period to be  $P_{\rm rot}$ = 144 - 306~days. 
 Using the theoretical predictions of tidal interaction and pseudosynchronization, 
 we estimate the orbital eccentricity  $e=0.68-0.82$.
 We briefly discuss the connection of our results with the photometric variability 
 of the object.
}
\maketitle

\section{Introduction} 
MWC~560 (V694~Mon) was discovered as an object with 
bright hydrogen lines (Merrill \& Burwell 1943). 
It is a symbiotic binary system,
which consists of a red giant and a white dwarf   
(Tomov et al. 1990; Michalitsianos et al. 1993).
The orbital period is supposed to be $P_{\rm orb} \approx 5.3$~yr  (Doro-shenko, Goranskij \& Efimov 1993).
The most spectacular features of this object are the collimated ejections 
of matter with velocities of up to $\sim 6000$~\kms\ (Tomov et al. 1992; Stute \& Sahai 2009)
and the resemblance of its emission line spectrum to that of 
the low-redshift quasars (Zamanov \& Marziani 2002).

The jet ejections are along the line of sight and the 
system is seen almost pole-on ($i < 16^\circ$).
This makes it difficult to obtain the orbital parameters using 
the radial velocity variations, and impossible to observe effects 
such as eclipse or illumination. 

To improve our understanding of this object, we aim here to: (1) measure the projected rotational velocity 
of the red giant (\vsi) from high resolution spectra, and (2) use our measured \vsi to find the eccentricity $e$ of the orbit, on the basis of the 
theory for pseudosynchronization in binary stars.

\section{Spectral Data and \vsi\  measurement}
\label{spec}

\subsection{Observations}

\begin{figure*}
 %\mbox{} 
 \vspace{7.0cm}  
 \includegraphics{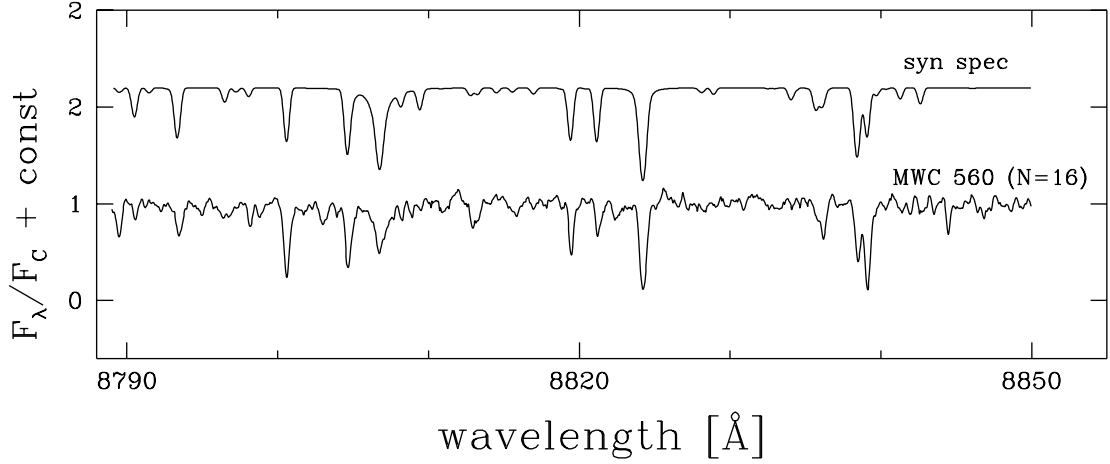}   
 \caption[]{The average spectrum of MWC~560 in the range $\lambda = 8790 - 8850$~\AA, obtained from 
 16 exposures with FEROS spectrograph at the 2.2-m telescope at La Silla, 
 and a synthetic spectrum using $T_{\rm eff}$=3424~K, $\log g=0.5$ $\xi=3$~\kms, \vsi = 8.2~\kms, 
 and 64\% contribution from the giant. }
    \label{MWC.spec1}
 \end{figure*}

We analyzed 21 high resolution spectra of MWC~560, obtained with the FEROS spectrograph.
FEROS is a fiber fed echelle spectrograph, providing high resolution of
$\lambda/\Delta \lambda = 48000$ and an wide wavelength coverage from about  
4000~\AA\  to 8900~\AA\ in one exposure (Kaufer et al. 1999). 
16 of the analyzed spectra are from our observations of the MWC~560 with 
the 2.2-m telescope at ESO, La Silla (ESO program 074.D-0114). 
The remaining spectra (5), are downloaded from FEROS Spectroscopic Database at the 
LSW Heidelberg\footnote{
www.lsw.uni-heidelberg.de/projects/instrumentation/Feros/ferosDB/} and were obtained when 
FEROS was mounted on the 1.52-m telescope at La Silla. A log of observations is given in 
Table~\ref{table1}. Details of the data processing are given 
in Zamanov et al. (2007). In Fig.~\ref{MWC.spec1}  we show the average spectrum derived from 
the 16 spectra obtained by 2.2-m telescope.

Symbiotic stars have composite spectra with three main sources of radiation --
red giant, hot component and nebula. Modeling of a few objects is given 
in Skopal (2005). 
Fig.~\ref{MWC.spec1} also shows a synthetic spectrum of 
a M5III star with parameters: $T_{\rm eff} =3424$~K, $\log g = 0.5$, $\xi=3$~\kms,
\vsi = 8.2~\kms, and 64\% contribution of the giant at $\lambda =8800$~\AA\, (see 2.2.).

%%-------------------------------------------------------------------------------
\begin{table}
  \begin{center}
  \caption{Spectral observations of MWC~560. The columns in the table represent: 
  (1) modified Julian Date (MJD) of the start of the exposure, 
  (2) exposure time, and the
  (3) measured projected rotational velocity (\vsi) of the mass donor. }
  \begin{tabular}{cccccc}
  
\hline 

   MJD   &  exp-time  &    \vsi  & \\
         &  [sec]     &    [\kms]       &        & \\
         &            &     \\
% \medskipFileName                object          MJD-OBS         exptime
\hline

\\
2.2-m telescope  \\
53314.279 & 300  &      8.5   \\
53314.283 & 300  &      8.6   \\
53314.287 & 300  &      7.9   \\
53314.291 & 300  &      7.8   \\
53314.295 & 300  &      8.0   \\
53314.299 & 300  &      8.2   \\
53314.303 & 300  &      8.2   \\
53314.308 & 300  &      8.5   \\
53314.312 & 300  &      8.8   \\ 
53314.316 & 300  &      7.8   \\
53314.320 & 300  &      8.3   \\
53314.324 & 300  &      7.7   \\
53314.328 & 300  &      8.6   \\
53314.332 & 300  &      7.9   \\
53314.336 & 300  &      8.1   \\
53314.340 & 300  &      8.5   \\  \\
1.52-m telescope \\
51133.223 & 1200 &     7.8\\
51131.248 & 2400 &     8.9\\
51135.283 & 1200 &     7.8\\
51136.259 &  900 &     7.9\\
51141.315 & 1200 &     8.2\\
\hline 
\end{tabular}
  \label{table1}
  \end{center}
\end{table}

\subsection{\vsi\  measurement}

We measure \vsi\ by comparing the FWHM of spectral lines of observed 
and synthetic spectra. The procedure is similar to that described in Fekel (1997).

We synthesized spectra by using the code SYNSPEC 
(Hubeny, Lanz \& Jeffery 1994) in the spectral region  $\lambda =$ 8750 - 8850~\AA. 
Atmospheric parameters  $T_{\rm eff}=3424$~K, $\log g=0.5$ (typical for a M5III star, see van Belle et al. 1999),
the instrumental profile for our FEROS spectra, and solar abundances
were the input parameters for the code.
LTE model atmospheres were extracted from Kurucz's grid (1993). 
The VALD atomic line database (Kupka et al. 1999) was used to create a line list 
for spectrum synthesis. 

We adopt a microturbulent velocity  $\xi=3$ \kms. 
This value is the same as accepted by  Fekel, Hinkle \& Joyce (2004)
and it is close to the values for other symbiotics.
Schmidt et al. (2006)  calculated $\xi=2.2$ \kms\ for CH~Cyg.
Wallerstein et al.(2008) used $\xi=2.2$~\kms\  for T~~CrB.
The relation between $\xi$ and $\log g$ (Gratton, Carretta, \& Castelli 1996)
gives $\xi \approx 2.1$~\kms\ for our case. 
% {\bf however we prefer 3~\kms because...}

A grid of synthetic spectra for projected rotational velocities from 0 \kms\ to 20 \kms\ 
%{\bf at a resolution of... \kms} 
was  calculated. 
\vsi \, was calculated by measuring the full width at half maximum (FWHM) of twelve observed spectral lines.
These were compared to the spectral line half-widths 
from the synthetic spectra. 
We thereby derive a mean \vsi $= 8.2$~\kms, and a 
standard deviation of the mean $\sigma = 0.4$~\kms using all 21 spectra. 
we note that if we adopt a microturbulent velocity $\xi=2$~\kms\ we 
would have measured \vsi $= 9.4$ \kms. 
For the error of our measurements we adopt a conservative value 
$\pm 1.5$~\kms, which also includes the uncertainty of $\xi=3 \pm 0.5$~\kms\ 
and the error of our method.
We therefore estimate our final measurement to be \vsi $= 8.2 \pm 1.5$~\kms. 

% For the error of our measurements we adopt a more conservative value:
% \vsi $= 8.2 \pm 1.0$~\kms, which also includes the error of our method.

%For the error of our measurements we adopt a more conservative value:
%\vsi $= 8.2 \pm 1.0$~\kms, which also includes the error of our method.
%In our measurement we adopt the microturbulent velocity $\xi=3$~\kms\ 
%(if $\xi=2$~\kms\ is assumed,  then  \vsi $= 9.4$ \kms).

% \bibitem[Schmidt et al.(2006)] Schmidt, M.R., Zacs L., Miko{\l}ajewska J., Hinkle, K.H.: 2006, A\&A~446, 603
% \bibitem[Gratton et al.(1996)]{1996A&A...314..191G} Gratton, R.~G., Carretta, E., \& Castelli, F.\ 1996, \aap, 314, 191 
% \bibitem[Wallerstein et al.(2008)]{2008PASP..120..492W} Wallerstein, G., Harrison, T., Munari, U., \& Vanture, A.\ 2008, \pasp, 120, 492 
% \bibitem[Fekel et al.(2004)]{2004IAUS..215..168F} Fekel, F.~C., Hinkle, K.~H., \& Joyce, R.~R.\ 2004, Stellar Rotation, 215, 168 
		       			                              
% \bibitem[Dyck et al.(1996)]{1996AJ....111.1705D} Dyck, H.~M., Benson, J.~A., van Belle, G.~T., \& Ridgway, S.~T.\ 1996, \aj, 111, 1705 

% \bibitem[van Belle et al.(1999)]{1999AJ....117..521V} van Belle, G.~T., et al.\ 1999, \aj, 117, 521 

% The symbiotic stars have composite spectra with three main sources of radiation --
% red giant, hot component and nebula. Detailed modeling for a few objects is given in Skopal (2005). 
% Here we find that the red giant in MWC~560 contributes $\sim 64$\% of the continuum at $\lambda 8800$~\AA .                                                   

%%%------------------------------------------------------------------------------  
%%%------------------------------------------------------------------------------- 

\section{Tidal interaction in MWC~560}

\subsection{Parameters of MWC~560}
% {\bf Iain suggests to move this subsection 3.1. between 1. Introduction and 2. Spectral data}

Meier et al. (1996) classified the cool component of MWC~560 as 
M5III-M6III giant. 
M{\"u}rset \& Schmid (1999) give M5.5III - M6III as its stellar type.
According to van Belle et al. (1999), a typical M5III giant has the following parameters: 
$R_{\rm g} = 139.6 \pm 5.6~$R$_\odot$, 
$T_{\rm eff}= 3424$~K, $L_{\rm g} \sim 2410~L_\odot$, 
and a typical M6III giant: $R_{\rm g} = 147.9 \pm 7.7~R_\odot$, $T_{\rm eff}= 3375$~K,
$L_{\rm g} \sim 2560~L_\odot$.
We adopt for the M giant in MWC~560 % to have
$R_{\rm g} = 140 \pm 7~R_\odot$ and $L_{\rm g} \sim 2400~L_\odot$. 

For the masses of the red giant and white dwarf we will assume $M_{\rm g} = 1.7~M_\odot$ and
% za MWC560 we calculate  $\log g = 0.372  [cm/s^2 ] !!!
$M_{\rm wd}=0.65~M_\odot$, respectively. These values are the average masses of the components of the 
symbiotic stars (Miko{\l}ajewska 2003). 
Concerning the orbital period, we use $P_{\rm orb}=1931 \pm 162$~day (Gromadzki et al. 2007). 
With the above values of the parameters assumed, we derive the semimajor axis of the orbit 
to be $a \approx 860~R_\odot$.

The jet (orbit) inclination angle to the line of sight
is $i < 16^\circ$, and the white dwarf accretes at a rate
$\dot M_{\rm acc} \approx 5 \times 10^{-7}~M_\odot$ (Schmid et al. 2001).

\subsection{Synchronization and circularization time scales}
\label{Sect.times}

The physics of tidal synchronization for stars with convective
envelopes has been analyzed several times.
There are some differences in the analysis of different authors, leading to varying
synchronization timescales. We use the estimate from Zahn (1977,
1989). The synchronization timescale in terms of the period is
\begin{equation}
 \tau_{\rm syn} \approx 800 \left( \frac{ M_{\rm g}  R_{\rm g}}{ L_{\rm g}}\right)^{1/3} 
 \frac{M_{\rm g}^2 (\frac{M_{\rm g}}{M_{\rm wd}} + 1)^2}{R_{\rm g}^6} P_{\rm orb}^4\ \;   {\rm yr},
\label{sync}
\end{equation}
where $M_{\rm g}$ and $M_{\rm wd}$ are the masses of the giant and white dwarf
respectively in Solar units, and $R_{\rm g}$, $L_{\rm g}$ are the radius and
luminosity of the giant, also in Solar units. The orbital period $P_{\rm orb}$ is
measured in days. 

Following Hurley, Tout \& Pols (2002), 
the circularization time scale is:
\begin{equation}
 \frac{1}{{\tau}_{\rm circ}} = \frac{21}{2} \left(\frac{k}{T}\right) q{_2} (1+q{_2}) 
 \left(\frac{R_{\rm g}}{a}\right)^8 .
\label{circ}
\end{equation}
where $q_2 $ is the mass ratio $q_2 = M_{\rm wd}/M_{\rm g}$,
In  Eq.~\ref{circ}, $(k/T)$ is 
derived from Rasio et al. (1996):
\begin{equation}
\left(\frac{k}{T}\right) = \frac{2}{21} \frac{f_{\rm conv}}{{\tau}_{\rm conv}} \frac{M_{\rm env}}{M_{\rm g}}  
\;  {\rm yr}^{-1} ,
\label{KT}
\end{equation}
where $R_{\rm env}$ is the depth of the convective envelope, $M_{\rm env}$ is the envelope's mass, and
\begin{equation}
{\tau}_{\rm conv} = 0.4311 \left(\frac{M_{\rm env}R_{\rm env}(R_{\rm g}-\frac{1}{2}R_{\rm env})}{3 L_{\rm g}}\right) ^{\frac{1}{3}}  {\rm yr}
\label{tau}
\end{equation}
is the eddy turnover time scale (the time scale on which the largest convective cells turnover). 
The numerical factor $f_{\rm conv}$ is
\begin{equation}
{f}_{\rm conv} = {\rm min} \left[1, \left( \frac{P_{\rm tid}}{2{\tau}_{\rm conv}} \right) ^2 \right],
\label{fconv}
\end{equation}
where $P_{\rm tid}$ is the tidal pumping time scale given by
\begin{equation}
\frac{1}{P_{\rm tid}} = \left|\frac{1}{P_{\rm orb}} - \frac{1}{P_{\rm rot}}\right|.
\label{ptid}
\end{equation}
For MWC~560  we assume for the red giant to have $R_{\rm env}=0.9~R_{\rm g}$ and $M_{\rm env}= 1.0~M_\odot$ (Herwig 2005).
Using these parameters, we calculate $P_{\rm tid}= 248$~days, 
$f_{\rm conv}=1$, $\tau_{\rm conv}=0.476$~yr, 
and $(k/T)= 0.12$ yr$^{-1}$. Then from Eq.~\ref{sync} and Eq.~\ref{circ} follows
that the synchronization and circularization time scales are
$\tau_{\rm sync} = 2.6 \times 10^4$~yr and 
$\tau_{\rm circ} = 3.1 \times 10^6$~yr respectively.

% The pseudosynchronization timescale is shorter then  for 

Following  Hut (1981), we estimate 
the  pseudosynchronization timescale $\tau_{\rm ps}$: 

\begin{equation}
  \tau_{\rm ps} = \frac {7}{3(\alpha - 3)} \tau_{\rm circ},
\end{equation}

where $\alpha$ is a dimensionless quantity, representing the ratio of the orbital and rotational angular
momentum:

\begin{equation}
\alpha = \frac {q_2}{1+q_2} \frac {1}{r_{\rm g}^{2}} \left( \frac {a}{R_{\rm g}} \right) ^2 ,
\end{equation}
where  $r_{\rm g}$ is the gyration radius of the giant.
For a red giant we adopt $r_{\rm g} \approx 0.3$ (Claret, 2004, 2007).
%Using these values, 
We calculate $\alpha = 800$ and 
$\tau_{\rm ps}= 9.1 \times 10^3$~yr.

% ratio   $\tau_{\rm ps} / \tau_{\rm circ} = 2.9 \times 10^{-3}$.

%It then follows from Eq.~\ref{sync} and Eq.~\ref{circ} 
%that the synchronization and circularization time scales are
%$\tau_{\rm sync} = 2.6 \times 10^4$~yr and 
%$\tau_{\rm circ} = 3.1 \times 10^6$~yr respectively.

%  
%
%  \begin{equation}
%  \tau_{\rm l} = \frac {7}{\alpha +1} \tau_{\rm circ},
%  \end{equation}
%
%   \tau_{\rm l}=2.7e4 = 2.7\times 10^4  yr

\subsection{Lifetime of the symbiotic phase}

% The lifetime of a reg giant with solar mass is $6\times10^8$~years (0.6~Gyr)
% (Sackmann, Boothroyd \& Kraemer  1993).  ?????
 
 The typical lifetime of a symbiotic star is $\tau_{\rm ss} \sim 10^5$~yr (Yungelson, et al. 1995;
 L{\"u}, Yungelson, \& Han 2006). 
 In the case of MWC~560, we can estimate from the rate of accretion on the white dwarf, 
 $\dot M_{\rm acc} \approx 5 \times 10^{-7}~M_\odot$ (Schmid et al. 2001), that it will take $10^6$~yr 
 to accrete $\sim 0.5~M_\odot$ from the envelope of the red giant companion. 
 Because the giant also losses mass via stellar wind, we find that the lifetime 
 of the symbiotic phase of MWC~560 should be 
 $\tau_{\rm ss} \simlt 10^6$~yr.

For MWC~560 we have therefore the situation in which $\tau_{\rm ps} < \tau_{\rm syn} < \tau_{\rm ss} < \tau_{\rm circ}$. 
This means that the symbiotic phase is long enough that the tidal forces can 
synchronize (pseudosynchronize)  the rotation of the red giant. 
On the other hand, the value of $\tau_{\rm circ}$ demonstrates that 
the symbiotic lifetime of MWC~560 is shorter than the circularization time, 
and therefore the orbit can be eccentric.
This is in agreement with the observational evidences found by 
Fekel et al. (2007) that the symbiotic stars with P$_{\rm orb} > 800$~days 
tend to have eccentric orbits.

Schmutz et al. (1994) estimated that the timescale for circularization in 
SY~Mus is $\sim 10$ times longer then the synchronization time. 
For MWC~560, we find even higher ratio:
 $\tau_{\rm circ} / \tau_{\rm syn} \approx 45$. 

The above implies that in MWC 560, the red giant is probably more or less synchronized,
but the orbit is not circularized.

\subsection{Rotation and pseudosynchronization}
In a binary with a circular orbit the rotational period of the primary, P$_{\rm rot}$, 
reaches an equilibrium value at the orbital period, $P_{\rm rot} = P_{\rm orb}$.  
However, in a binary with an eccentric orbit, the 
tidal force acts to synchronize the rotation of the mass donor with 
the motion of the compact object at the periastron -- the effect called 
pseudosynchronous rotation (Hall 1986).
The corresponding equilibrium (i.e. pseudosynchronization) is reached 
at a value of $P_{\rm rot}$ which is
{\it less} than $P_{\rm orb}$, the amount less being a function of the orbital
eccentricity $e$. 
Hut (1981) showed that the period of pseudosynchronization, P$_{\rm ps}$, is :

\begin{equation}
P_{\rm ps} = \frac{(1+3e^2+\frac{3}{8}e^4)(1-e^2)^\frac{3}{2}}{1+\frac{15}{2}e^2+
\frac{45}{8}e^4+\frac{5}{16}e^6} P_{\rm orb}.
\label{Eq-ps}
\end{equation}
When the eccentricity tends to zero, P$_{\rm ps}$ tends to P$_{\rm orb}$.

% At low eccentricity of the orbit  $e \rightarrow 0$ and P$_{\rm ps} \rightarrow  P_{\rm orb}$.

\subsection{Orbital eccentricity of MWC~560}

In order to determine the orbital eccentricity of  MWC~560, we first need 
to calculate $P_{\rm rot}$ for the mass donor. We use
\begin{equation}
P_{\rm rot}=\frac{2\pi R_{\rm g} \sin i}{v\: \sin i}. 
\label{Eq-Prot}
\end{equation}
The underlying assumption is that the rotational axis of the mass donor is 
perpendicular to the orbital plane.

Using the value for \vsi $= 8.2 \pm 1.5$~\kms\  (Sect.\ref{spec}), 
$R_{\rm g} = 140 \pm 7~R_\odot$, and  $\; i=12^\circ-16^\circ$ (Schmid et al. 2001),  
we calculate $P_{\rm rot}=144 - 306$~days. This value 
is considerably less than the orbital period, $P_{\rm orb}=1931 \pm 162$~days.
Following the results in Sect.\ref{Sect.times}, MWC~560 should 
be close to synchronization or pseudosynchronization, 
and $P_{\rm rot} =P_{\rm ps}$. 
Finally, using Eq.~\ref{Eq-ps} we can therefore estimate the orbital eccentricity to be 
$e=0.68-0.82$.

%This implies that in MWC~560, the rotation of the  red giant
% is  pseudosynchronized with the orbital motion of the white dwarf, 
% but the orbit is not circularized in agreement with our 
% timescale estimations. 

%---------------------------------------------------------------------

\section{Discussion}

In most of the symbiotics the activity of the hot components 
is irregular/aperiodic. However, in MWC~560 it has a periodic character. 
An analysis of the historical light curve (Luthardt 1991) 
and B-band photometry, 
as well as the periodogram analysis of V band and near-IR 
observations reveal that this periodicity has remained in-phase for over a century 
(Doroshenko et al. 1993; Gromadzki et al. 2007).
The most natural explanation for this phenomenon is orbital modulation: 
around periastron, the mass accretion rate increases, which causes
an increase of the accretion luminosity and the brightness of  
the accretion disk, which produces the observed modulation of the light curves  
(see also Doroshenko et al. 1993; Gromadzki et al. 2007).

%The estimated timescales of tidal interaction, and the finding that the orbit is 
%highly eccentric give one more evidence in this direction. 

With the parameters assumed above, we can calculate the distance between the components 
at the periastron: $r_{\rm p}=a(1-e) = 180 - 230~R_\odot$. It follows that the red giant
fills the Roche lobe for about  5\% -- 25 \% of the orbital period.  
This is in agreement with the suppositions of Gromadzki et al. (2007),
that the orbital eccentricity and corresponding Roche lobe overflow at the periastron
is the reason for the observed photometric variability. 
% A giant filling the Roche lobe, will lose mass at a rate $\sim 10^{-6}$ ???
% average mass accretion rate .....

It has been noted that the rotational period of the mass donor
is considerably shorter than the orbital period (Zamanov et al. 2008) in 
a few jet-ejecting symbiotics. 
Our findings here pose the question whether 
their orbits are also eccentric.

\section*{Conclusions}
In this note, we presented new measurements of the projected rotational velocity
of the red giant in the symbiotic star MWC~560. We find that \vsi $= 8.2 \pm 1.5 $~\kms.
On the basis of the theory of tidal interaction in binaries, we calculate that the orbit 
should be highly eccentric, with $e$\simgt 0.7. Our findings support the model that the observed photometric variability of MWC~560
is connected with high orbital eccentricity and Roche lobe overflow at periastron.

\acknowledgements
This work was supported by Bulgarian NSF (HTC01-152)
and Slovenian Research Agency (BI-BG/09-10-006).
We are grateful for useful comments from an anonymous referee, 
and Prof. I.A. Steele for English language corrections.

\end{document}